\documentstyle[aps,epsfig,times,multicol,color]{revtex}
\newcommand{\Label}[1]{\label{#1}}
\newcommand{\BEC}{Bose-Einstein condensate}
\newcommand{\GPE}{Gross-Pitaevskii equation}

\newcommand{\pdfstyle}{
\newcommand{\Endrule}{\vskip 3pt\noindent\hrule width 8.6cm\vskip 3pt}
\newcommand{\Beginrule}{\vskip 3pt\noindent\hbox{%
\vbox{\hbox to 9cm{\hfill}}\vbox{\hrule width 9cm}} \vskip 3pt}
\newcommand{\StartTwoColumn}{\begin{multicols}{2}}
\newcommand{\EndTwoColumn}{\end{multicols}}
\renewcommand{\narrowtext}{\Beginrule\begin{multicols}{2}}
\renewcommand{\widetext}{\end{multicols}\Endrule}
\newcommand{\NP}{}}

\def\DRAFT{%
\renewcommand{\Label}[1]{\label{##1}%\message{##1}%
{\hbox to 0cm{\textcolor{green}{\hss\em ##1\quad}}}}
\def\Input##1{\include{##1}}}

\pdfstyle

\newcommand{\muc}{\mu_{\rm C}}
\newcommand{\muinit}{\mu_{\rm init}}
\newcommand{\mueff}{\mu_{\rm eff}}

\begin{document}
\title{Quantum kinetic theory VII: The influence of 
vapor dynamics on condensate growth}
\author{M. J. Davis$^1$, C. W. Gardiner$^2$, and R. J. Ballagh$^3$}
\address{$^1$Clarendon Laboratory, Department of Physics, University of Oxford,
 United Kingdom}
\address{$^2$School of Chemical and Physical Sciences, Victoria University of
 Wellington, New Zealand}
\address{$^3$Physics Department, University of Otago, Dunedin, New Zealand} 
\maketitle
\begin{center}
\today
\end{center}
\begin{abstract}

We extend earlier models of the growth of a Bose-Einstein condensate 
\cite{BosGro1,newerBosGro,BosGro2} to include the full dynamical  effects of
the thermal cloud by  numerically solving a  modified quantum Boltzmann
equation.  We determine the regime in which  the assumptions of the simple
model of \cite{BosGro2} are a reasonable approximation,  and compare our new
results with those that were earlier compared with  experimental data.   We
find good agreement with our earlier modelling, except at higher  condensate
fractions, for which a significant speedup is found. We also investigate the
effect of the final temperature  on condensate growth, and find that this has a
surprisingly small  effect.

The particular discrepancy between theory and experiment found in our earlier
model remains, since the  speedup found in these computations does not occur in
the parameter regime specified in  the experiment.

\end{abstract}


\StartTwoColumn
\section{Introduction}\Label{sec1}

The fundamental process in the growth of a \BEC\ is that of 
{\em bosonic stimulation}, by which atoms are scattered into and out 
of the condensate at rates enhanced by a factor proportional to the number of 
atoms in the condensate.  This was first quantitatively considered by 
Gardiner {\em et al.} \cite{BosGro1}, in a paper which treated the 
idealized case of the growth of a condensate from an nondepletable 
``bath'' of atoms at a fixed positive chemical potential $\mu$ and 
temperature $T$.  This gave rise to a simple and elegant formula known 
as the {\em simple growth equation}
\begin{eqnarray}
\dot{n}_0 = 2W^+(n_0)\left\{\left(1-\mbox{e}^{[\muc(n_0)-\mu]/kT}\right)n_0 + 1
\right\},
\label{eqn:simple_growth}
\end{eqnarray}
in which $n_0$ is the population of the condensate, $\mu$ is the 
chemical potential of the thermal cloud, and $\muc(n_0)$ is the 
condensate eigenvalue.  The prefactor $W^+(n_0)$ is a rate with an 
expression derived from quantum kinetic theory \cite{QK}, which was 
estimated approximately in \cite{BosGro1} by using a classical 
Boltzmann distribution.  To go beyond the Boltzmann approximation for 
$W^+$ involves a very much more detailed treatment of the populations of 
the trap levels with energy less than $\mu$, since the equilibrium 
Bose-Einstein distribution for $\mu > 0$ is not consistent with 
energies less than $\mu$.  In other words, the populations of the 
lower trap levels {\em cannot} be treated as time-independent, and thus 
the dynamics of growth must include at least this range of trap levels 
as well as the condensate level.  Therefore in 
\cite{newerBosGro,BosGro2} we considered a less simplified model, 
covering a range of energies up to a cut-off, $E_R$, above which the 
system was assumed to be a thermal cloud with a fixed temperature and 
chemical potential.  Equations were derived for the rate of growth of 
these levels along with the condensate, and the rates at which 
particles from the thermal bath scattered these quasi-particles 
between levels within the condensate band.  
The results of calculations showed that a speedup of the growth rate 
by a factor of the order of 3--4 compared to the simple growth 
equation could be expected, and that the initial part of the growth 
curve would be modified, leading to a much sharper onset of the 
initiation of the condensate growth.

The only experiment that has been done on condensate growth 
\cite{MITgrowth} was then under way.  In these experiments clouds of 
sodium atoms were cooled to just above the transition temperature, at 
which point the high energy tail of the distribution was rapidly 
removed by an very severe RF ``cut'', where the frequency of the RF 
field was quickly ramped down.  After a short period of equilibration, 
the resulting vapor distribution was found to be similar to the 
assumptions of our theoretical treatments, and condensate growth 
followed promptly.  The results obtained were fitted to solutions of 
the simple growth equation (\ref{eqn:simple_growth}).  When 
experimental results 
became available a speedup of about the predicted factor was found, 
and indeed the higher temperature results agreed very well with the 
theoretical predictions.  At lower temperatures there was still some 
disparity; the theory predicted a slower rate of growth with 
decreasing temperature, but experimentally the opposite was observed.

The situation in which we now find ourselves leaves no alternative 
other than to address the remaining approximations.  In our previous 
work we have made four major approximations:
\begin{itemize}
\item[(i)]  The part of the vapor with energies higher than $E_R$ has 
been treated as being time independent.

\item[(ii)]  The energy levels above the condensate level were modified 
phenomenologically to account for the fact that they must always be 
greater than the condensate chemical potential, which rises as the 
condensate grows.

\item[(iii)] We treated all levels as being particle-like, on the 
grounds that detailed calculations \cite{StringariRev} have shown that
only a very small proportion of excitations of a trapped Bose gas are not of 
this kind.

\item[(iv)] We have used the quantum Boltzmann equation in a ergodic 
form, in which all levels of a similar energy are assumed to be 
equally occupied.

\end{itemize} In this paper we will no longer require the first two of these 
approximations.  Abandoning the first means that we are required to  take care
of all kinds of collisions which can occur, and thus treat the time-dependence
of all levels.  This comes at a dramatic increase in  both the computation time
required (hours rather than seconds) and the  precision of algorithms
required.  We also use a density of states  which should be close to the actual
density of states as the  condensate grows, thereby avoiding the
phenomenological modification  of energy levels.  However, we still treat all
of the levels as being  particle-like, since it seems unlikely that the few
non-particle-like  excitations will have a significant effect on the growth as
a whole. The ergodic form of the quantum Boltzmann is needed to make the 
computations tractable, and is of necessity retained.

\section{Formalism}\Label{sec2}
The basis of our method is Quantum Kinetic theory, a full exposition of which
is given in Ref.~\cite{QK}.  These develop a complete framework for the study
of a trapped Bose gas in  a set of master equations.   The full solution of
these equations is not feasible, however,  and therefore some type of
approximation must be made.  The basic structure of the method used 
here is essentially the same as that of QKVI, the major difference 
being that all time-dependence of the distribution function is 
retained.  As explained in  \cite{BosGro2,QKVI}, quantum kinetic 
theory leads to a model which can be viewed as a modification of the 
quantum Boltzmann equation in which

\begin{itemize}
\item[i)] The condensate wavefunction and energy eigenvalue---the 
condensate chemical potential $\muc(n_0)$---are given by the 
solution of the time-independent \GPE\ with $n_0$ atoms.

\item[ii)] The trap levels above the condensate level are the quasiparticle 
levels 
appropriate to the condensate wavefunction.  This leads to a density 
of states for the trap levels which is substantially modified, as 
discussed below in Sect.~\ref{density}

\item[iii)]  The transfer of atoms between levels is given by a 
modified quantum Boltzmann equation (MQBE) in the
energy representation.  This makes the ergodic assumption; that the
distribution function depends only on energy. 
\end{itemize}

%%%%%%%%%%%%%%%%%%%%%%%%%%%%%%%%%%%%%%%%%%%%%%%%%%%%%%%%%%%%%%%%%%%%%%%%%
\subsection{The ergodic form of the quantum Boltzmann equation}   
\label{derivation}
The derivation of the ergodic form of the quantum Boltzmann equation 
used by \cite{HollandKE} is particular to the undeformed harmonic 
potential, and we give here a derivation appropriate to our case, in 
which the density of states can change with time as the condensate 
grows.  We {\em bin} the phase space into energy bands labeled by the 
index $n$ with energies in a range 
$D_n(t)\equiv\left(\varepsilon_n(t)-{\delta\varepsilon_n(t)\over2},
\varepsilon_n(t)+{\delta\varepsilon_n(t)\over2} 
\right)$ of width $\delta\varepsilon_n(t)$, and these widths change in time so 
that the number of states within each bin, $ g_n$ is constant in time.

Starting from the full quantum Boltzmann equation the ergodic approximation is 
expressed in terms of this binned description as follows:  We set
$ f({\bf x},{\bf K},t)  $ equal to a constant, $ f_n$, when 
$ \epsilon({\bf x},{\bf K},t)\equiv
{ \hbar^2{\bf K}^2/ 2m} + V_{\rm eff}({\bf x},t)$ is inside the $ n$th bin, 
i.e., $\epsilon({\bf x},{\bf K},t) \in D_n(t)$.  
(Here $V_{\rm eff}({\bf x},t)$ is the potential of the trap, as 
modified by the mean field arising from the presence of the 
condensate, as explained in QKV and QKVI.)

Thus we can approximate 
\begin{eqnarray}\Label{ergodic1}
{\partial  f({\bf x},{\bf K},t) \over\partial t} \to 
{\partial f_n \over\partial t}
 \mbox{\qquad if\qquad}
\epsilon({\bf x},{\bf K},t) \in D_n(t),
\end{eqnarray}
In order to derive the ergodic quantum Boltzmann equation, 
we define the indicator function $ \chi_n({\bf x},{\bf K},t)$ of the $ n$th bin 
$ D_n(t)$ by
\begin{eqnarray}\Label{ergodic2}
 \chi_n({\bf x},{\bf K},t) &=&1  \mbox{\qquad if\qquad}
\epsilon({\bf x},{\bf K},t) \in D_n(t),
\\
&=& 0 \mbox{\qquad otherwise.}
\end{eqnarray}
The number of states in the bin $ n $ will be given by  
$ g_n = \int d^3{\bf x}d^3{\bf K}\, \chi_n({\bf x},{\bf K},t)/h^3$, and
is held fixed.

The formal statement of the binned approximation is 
\begin{eqnarray}\Label{ergodic201}
f({\bf x},{\bf K},t) \to \sum_nf_n\chi_n({\bf x},{\bf K},t),
\end{eqnarray}
and the ergodic quantum Boltzmann equation is derived by substituting 
(\ref{ergodic201}) into the various parts of the quantum Boltzmann equation as 
follows.  For the time derivative part we make this replacement, and project 
onto $ D_n(t)$, getting
\begin{eqnarray}\Label{ergodic3}
\int {d^3{\bf x}d^3{\bf K}\over h^3} \chi_n({\bf x},{\bf K},t)
{\partial f({\bf x},{\bf K},t) \over\partial t}
&\to & g_n {\partial f_n\over\partial t}.
\end{eqnarray}
[Note that the expansion (\ref{ergodic201}) would mean that  delta function 
singularities at the upper and lower boundaries of $ D_n(t)$ would arise by 
differentiating $ f({\bf x},{\bf K},t) $ as defined in (\ref{ergodic201}), but 
the condition that $ g_n $ be fixed means that these are of equal and opposite 
weight, and cancel when integrated over $ D_n(t)$, givng a result consistent 
with (\ref{ergodic3}).]
We now replace $ {\partial f({\bf x},{\bf K},t) /\partial t} $ on the left hand 
side of (\ref{ergodic3}) by the collison 
integral that appears on the right hand side of the quantum Boltzmann equation, 
and substitute for $ f({\bf x},{\bf K},t)$ in the collision integral using 
(\ref{ergodic201}). [The streaming terms give no contribution, since the form
(\ref{ergodic201}) is a function of the energy $ \epsilon({\bf x},{\bf K},t)$.]

This leads to the {\em ergodic quantum Boltzmann equation} in the form
\widetext
\begin{eqnarray}\Label{ergodic4}
 g_n {\partial f_n\over\partial t} &=&{4a^2h^3\over m^2}
\sum_{pqr}\left\{f_pf_q(1+f_r)(1+f_n)-(1+f_p)(1+f_q)f_rf_n\right\}
%\nonumber\\ &&\qquad\times
\int {d^3{\bf x}d^3{\bf K}\over h^3} 
\int {d^3{\bf K}_1} 
\int {d^3{\bf K}_2} 
\int {d^3{\bf K}_3} 
\nonumber\\ &&\times
 \chi_n({\bf x},{\bf K}_1,t)
 \chi_p({\bf x},{\bf K}_2,t)
 \chi_q({\bf x},{\bf K}_3,t)
 \chi_r({\bf x},{\bf K},t)
%\nonumber\\ &&\qquad\times
\delta\left({\bf K}_1+{\bf K}_2-{\bf K}_3-{\bf K}\right)
\nonumber\\ &&\times
\delta\left(\epsilon({\bf x},{\bf K}_1,t) + \epsilon({\bf x},{\bf K}_2,t)
-\epsilon({\bf x},{\bf K}_3,t) - \epsilon({\bf x},{\bf K},t)\right).
\end{eqnarray}
The final integral is now approximated by the  method of\cite{HollandKE}
to give [$ a$ and $ \bar{\omega}$ are defined in Sect.~\ref{sec3}]
\begin{eqnarray}\Label{ergodic5}
 g_n {\partial f_n\over\partial t} &=&
{8ma^2\bar{\omega}^2\over\pi\hbar}
\sum_{pqr}\left\{f_pf_q(1+f_r)(1+f_n)-(1+f_p)(1+f_q)f_rf_n\right\}
{M(p,q,r,n)}\Delta(p,q,r,n).
\end{eqnarray}
Here $ \Delta(p,q,r,n) $ is a function which expresses the overall 
energy conservation, and is defined by
\begin{eqnarray}\Label{ergodic6}
\Delta(p,q,r,n) &=& 1 
\mbox{ when } |\varepsilon_p+\varepsilon_q-\varepsilon_r-\varepsilon_n| \le 
{\left |\delta\varepsilon_p +\delta\varepsilon_q 
+\delta\varepsilon_r +\delta\varepsilon_n \right |\over 2},
\\
&=& 0 \mbox{ otherwise}.
\end{eqnarray}
\narrowtext\noindent
Because we approximate $f({\bf x},{\bf K},t)$ by a constant value
within each $D_n(t)$, energy conservation means that
$\bar E \equiv \sum_n\varepsilon_n g_n f_n(t)$ is constant.  This 
follows from energy conservation in the full quantum Boltzmann 
equation, which also implies that
\begin{eqnarray}\Label{ergodic601}
&&\sum_{rn}\Delta(p,q,r,n)M(p,q,r,n)(\varepsilon_r+\varepsilon_n)
\nonumber \\ &&\qquad
=
(\varepsilon_p+\varepsilon_q)\sum_{rn}\Delta(p,q,r,n)M(p,q,r,n)
\end{eqnarray}
This is the limit to which the binning procedure defines energy conservation.

\NP\begin{figure}\centering
\epsfig{file=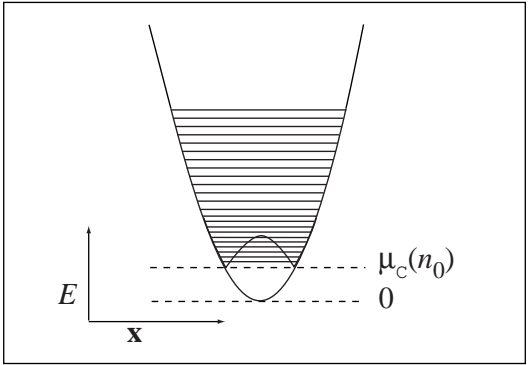,width=7cm}
\caption{Qualitative picture of the compression of the quantum levels above the
condensate mode as the condensate eigenvalue increases.} 
\label{fig:squash}
\end{figure}\NP

\section{Details of model}\Label{sec3}

The most important aspect of our model is the inclusion of the mean field
effects of the condensate.  As the population of the condensate increases, the
absolute energy of the condensate level also rises due to the atomic
interactions.  This results in a compression in energy space of the quantum
levels immediately above the condensate, (see Fig.~\ref{fig:squash}) and has an
important effect on the evolution of the cloud.

The correct description of the quantum levels immediately above the ground
state when there is a significant condensate population requires a
quasiparticle transformation.  This is computationally difficult, however, so
we make use of a single-particle approximation for these states.  This should
be reasonable, as most of the growth dynamics will involve higher lying states
that will be almost unaffected by the presence of the condensate.  In
\cite{BosGro2} we did this using a linear interpolation of the density of
states; here we use an  approximate treatment based on the 
Thomas-Fermi approximation.

\subsection{Condensate chemical potential $\muc(n_0)$}
We consider a harmonic trap with a 
geometric mean frequency of 
\begin{eqnarray}
\label{omegabar}
\bar{\omega}& =& (\omega_x \omega_y \omega_z)^{1/3}.
\end{eqnarray} 
We include the mean field effects via a Thomas-Fermi approximation for 
the condensate eigenvalue, which is directly related to the number of 
atoms in the condensate mode.  As in \cite{BosGro2,QKVI}, we use a 
modified form of this relation in order to give a smooth transition to 
the correct harmonic oscillator value when the condensate number is 
small:
\begin{eqnarray}
\muc(n_0) = \alpha \left[ n_0 + 
({3 \hbar \bar{\omega}}/{2 \alpha} )^{5/2} \right]^{2/5},
\label{eqn:mu_N}
\end{eqnarray}
where $\alpha = (15a \bar{\omega} m^{1/2}\hbar^2/4\sqrt{2})^{2/5}$, and $a$ is
the atomic \emph{s}-wave scattering length.  Thus, 
for $n_0 = 0$ we have $\muc(0) = \varepsilon_0 = 3\hbar \bar{\omega}/2$.

\subsection{Density of states $\bar g(\varepsilon)$}
\label{density}%
We assume a single particle energy spectrum with a Bogoliubov-like dispersion
relation, as in Timmermans {\em et al.} \cite{Timmermans}, which 
leads to a density of states of the form  
\widetext
\begin{eqnarray}
\bar g(\varepsilon,n_0) &=& \frac{4}{\pi} \frac{\muc(n_0)^2}{(\hbar \bar{
\omega})^3}
\left[
\left(\frac{\varepsilon}{\muc(n_0)} - 1\right)
 \int_0^1 dx \sqrt{1-x}\frac{
\left[\sqrt{(\varepsilon/\muc(n_0) - 1)^2 + x^2} - x \right]^{1/2}}{
\sqrt{(\varepsilon/\muc(n_0) -1)^2 + x^2}} + \int_1^{\varepsilon/\muc(n_0)} dx
\sqrt{x}
\sqrt{{\varepsilon\over\muc(n_0) }- x} \right],
\label{eqn:g_int}
\nonumber\\
\end{eqnarray}
This integral can be carried out analytically; the result is
\begin{eqnarray}
\bar g(\varepsilon,n_0)& =& \frac{\varepsilon^2}{2 (\hbar\bar{\omega})^3}
\left\{ 1 + q_1 \left( \muc(n_0)/\varepsilon \right) + 
\left(1-\frac{\muc(n_0)}{\varepsilon}\right)^2 q_2\left(\frac{1}{\varepsilon/
\muc(n_0) 
- 1}\right)\right\},
\label{eqn:d_o_s}
\end{eqnarray}
where 
\begin{eqnarray}
q_1(x) &=& \frac{2}{\pi}\left[\sqrt{x}\sqrt{1-x}(1-2x) -
 \sin^{-1}(\sqrt{x})\right], \\
q_2(x) & = & \frac{4 \sqrt{2}}{\pi} \left[
\sqrt{2x} +x \log\left(\frac{1+x+\sqrt{2x}}{\sqrt{1 + x^2}} \right)
-\left\{ \frac{\pi}{2} + \sin^{-1}\left(\frac{x-1}{\sqrt{1+x^2}}\right)
 \right\} \right]. 
\end{eqnarray}
\narrowtext

\begin{figure}\centering
\epsfig{file=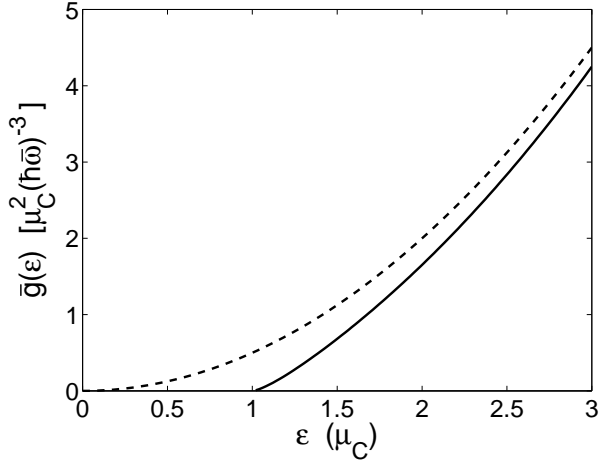,width=8cm}
\caption{The modified density of states (solid curve) compared with
non-interacting function (dashed curve) for the harmonic trap.} 
\label{fig:d_o_s}
\end{figure}

\noindent
This is plotted in Fig.~\ref{fig:d_o_s}, along with that for the ideal
gas.  Thus the density of states of the trap varies smoothly as the
condensate grows. 

\NP\NP

\section{Numerical methods} 

\subsection{Representation}
\begin{figure}\centering 
\epsfig{file=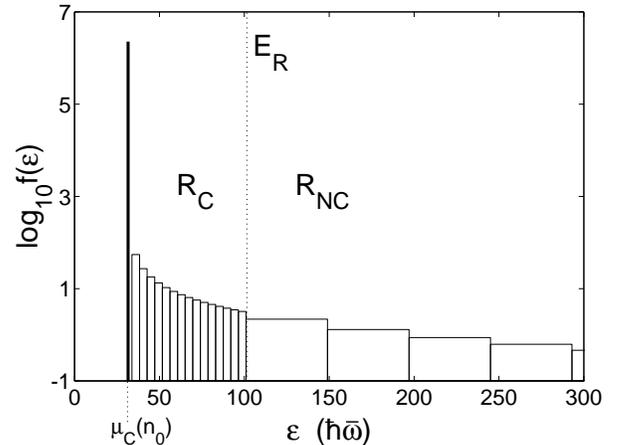,width=8cm} 
\caption{The numerical representation of the system with a condensate 
of $2.3 \times 10^6$ atoms at a temperature of $590$nK. $R_C$ is the 
condensate band, which is fine-grained, whereas $R_{NC}$ is the 
non-condensate band, which is coarse-grained.  The division between 
the two bands is fixed at $E_R$.  The condensate energy is derived 
from the Thomas-Fermi approximation.}
\label{fig:description}
\end{figure}

\Label{representation}
The bins we shall choose for the representation of the distribution in terms of 
the quantities $ f_n $ as in (\ref{ergodic201}) are
divided into two distinct regions, as shown 
diagrammatically in Fig.~\ref{fig:description}.  The lowest energy 
region corresponds essentially to the {\em condensate band $R_C$} of 
\cite{BosGro1,BosGro2,QKIII,QKVI}.  This is the region in which 
$f_n$ is rapidly varying in the regime of quantum 
degeneracy, and is described by a series of fine-grained energy bins  up
to an energy $E_R \approx 3\muc (n_{0,{\rm max}})$.  The condensate is a 
\emph{single} quantum state represented by the lowest energy bin.

\NP\NP

As the number of particles in the condensate changes, the energy of the
condensate level changes according to the Thomas-Fermi approximation of
Eq.~\ref{eqn:mu_N}.  Thus the total energy width of $R_{C}$ decreases as the
condensate grows.

We represent $R_{C}$ by a fixed number of energy bins of equal width $\delta
\varepsilon_n$ with a midpoint of $\varepsilon_n$ .

As the condensate energy
increases, we adjust $\varepsilon_n$ and $\delta \varepsilon_n$
{\em between} integration timesteps, such that the widths
of the bins remain equal. This is done by redistributing the numbers 
of particles into new bins after each timestep, and thus 
does not contradict the requirement that $g_n$ is fixed {\em during} 
the timestep.
We find that this is the most simple procedure
for the calculation of rates in and out of these levels.  We choose the number
of bins to be sufficient such that the width is not more than about $\delta
\varepsilon_n \sim 5 \hbar \bar{\omega}$.

The high energy region corresponds to the {\em thermal bath} of our 
previous papers.  This is the region in which $f_n$ is slowly varying, 
and therefore the energy bins are considerably broader (up to $64 
\hbar \bar{\omega}$ in the results presented in this paper).  The 
evaporative cooling is carried out by the sudden removal of population 
of the bins in this region with $\varepsilon_n > \varepsilon_{\rm 
cut}$.

\subsection{Solution}

\begin{figure}\centering
\epsfig{file=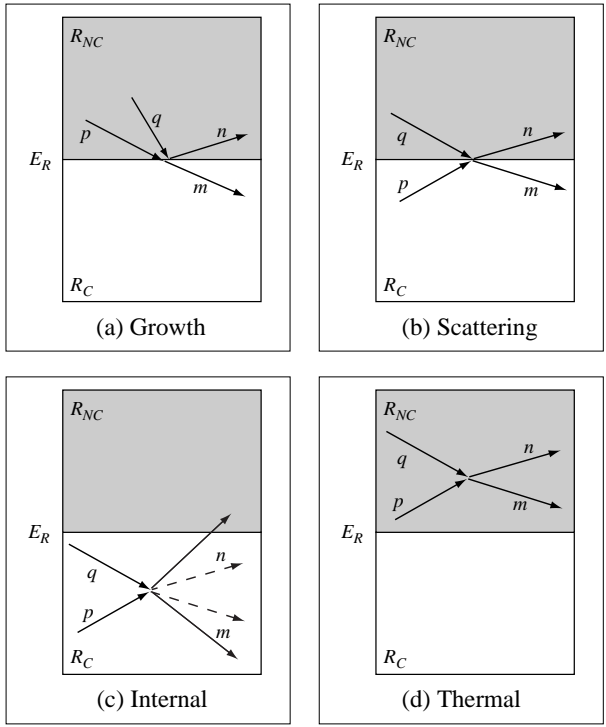,width=8cm}
\vskip 3mm
\caption{The four different collision types that can occur in our numerical
description.}
\label{fig:processes}
\end{figure}

There are four different types of collision that can occur given our numerical
description of the system.  These are depicted in Fig.~\ref{fig:processes}.

\NP\NP

\begin{itemize}

\item[(a)] {\em Growth}: This involves two particles in $R_{NC}$ 
colliding,
resulting in the transfer of one of the particles to the condensate band (along
with the reverse process).  

\item[(b)] {\em Scattering}: A particle in $R_{NC}$
collides with a particle in the condensate band, with one particle remaining 
in
$R_C$.  

\item[(c)] {\em Internal}: Two particles within the condensate band
collide with at least one of these particles remaining in $R_{NC}$ 
after the collision. 

\item[(d)] {\em Thermal}: This involves all particles involved in the collision 
coming
from the non-condensate band and remaining there.
\end{itemize}
Our first description of condensate growth \cite{BosGro1} considered 
only process (a).  The next calculation \cite{newerBosGro,BosGro2} involved 
both 
process (a) and (b).  The calculations presented below include all 
four processes, allowing us to determine whether the earlier 
approximations were justified.

The computation of the rates of processes (a) and (b) is made difficult 
because of the different energy scales of the two regions of the 
distribution function.  Our solution  is to  \emph{interpolate}
the distribution function $f_n$ in $R_{NC}$ (non-condensate
band) such that the bin sizes are reduced to be the same as for $R_{C}$
(the condensate band).  The rates are then calculated using this
interpolated distribution function, now consisting of more than one
thousand bins,  and the rates for the large bins of the non-condensate
band are found by summing the rates of the appropriate interpolated
bins.

We have found that these rates are {\em extremely} sensitive to the
accuracy of the numerical interpolation --- small errors lead to
inconsistencies in the solutions of the MQBE.  This procedure is
more efficient than simply using the same bin size for the whole
distribution, as there are only a small number of bins for the
condensate band. 

\subsection{Algorithm}
The algorithm we use to solve the MQBE is summarised as follows

\begin{itemize} 

\item[(1)] Calculate the collision summation for all types of
collisions, keeping the density of states, and the energies of the
levels in the condensate band $R_{C}$ fixed.  The distribution
function $f_n(t + \delta t)$ is calculated using an
embedded 4th order Runge-Kutta method, using Cash-Karp parameters
\cite{numerical_recipes}.

\item[(2)] 
The quantity $ {M(p,q,r,n)}$ defined by (\ref{ergodic5}) expresses all the 
overlap integrals, and is quite difficult to compute exactly.
In our 
computations we have simply set this to correspond to the value found in 
\cite{HollandKE}, i.e., we set
\begin{eqnarray}\Label{ergodic7}
M(p,q,r,n) &=& g_{{\rm min}(p,q,r,n)},
\end{eqnarray}
and express energy conservation in a simplified form,  using the 
fact that the energy bins will be chosen {\em equally spaced}, by 
choosing a Kronecker delta form
\begin{eqnarray}\Label{ergodic701}
\Delta(p,q,r,n) &\to & \delta(p+q,r+n).
\end{eqnarray}
The difference between these two forms clearly goes to zero as the 
bins become very narrow. 

It has been explicitly checked that in practice energy is conserved to a very 
high degree of accuracy throughout the calculation.

\item[(3)] As a result of the time step, the condensate population 
will have changed.  This causes the density of states to alter 
slightly, along with the positions and widths of the energy bins in 
the condensate band, as all these quantities are determined by the 
condensate number $n_0$.  The derivation in Sect.~\ref{derivation} 
shows that the populations $g_n f_n$ are of the bins which move with the 
change of energy levels and density of states as the condensate grows 
so as to maintain the number of levels $g_n$ in the bin constant.  
Therefore after the Runge-Kutta timestep, the numbers $g_n f_n$ represent 
the numbers of particles in the bins determined by the appropriate 
energy levels after that step.

\item[(4)] As a result of the preceding step, the bins will no longer of equal 
width, so we rebin the numbers of atoms into a new set of equally spaced bins, 
as explained Sect.~\ref{representation}.

To ensure total number conservation of particles, we keep the 
\emph{number} of particles in each bin, $g_n f_n$ 
constant when we adjust the energies and widths of the bins.  As the 
change in the density of states and the width of each bin is 
determined by the condensate number, the {\em occupation per energy 
level} of the $n$th bin, $f_n$, must be altered slightly 
to ensure number conservation.

\item[(5)] We now continue with step (1).

\end{itemize}

The change in $\mu_C(n_0)$ with each time step, and hence the shifts in the
energy of the bins in $R_C$  is very small.   Hence, the adjustment of the
distribution function due to step (3) is tiny, much smaller than the change due
to step (1).

The method has been tested by altering the position of $E_R$ and width 
of the energy bins of $R_{NC}$ and $R_{C}$.  We have found that the 
solution is independent of the value of $E_R$ over a large range of 
values of these parameters.

\section{Results}

In this paper we present the results of simulations modelling the 
experiments described in \cite{MITgrowth}.  In these 
experiments a cloud of sodium atoms confined in a `cigar' shaped 
magnetic trap was evaporatively cooled to just above the Bose-Einstein 
transition temperature.  Then, in a period of 10ms the high energy 
tail of the distribution was removed with a very rapid and rather 
severe RF cut.  The condensate was then manifested by the 
formation of a sharp peak in the density distribution.

\begin{figure}\centering
\epsfig{file=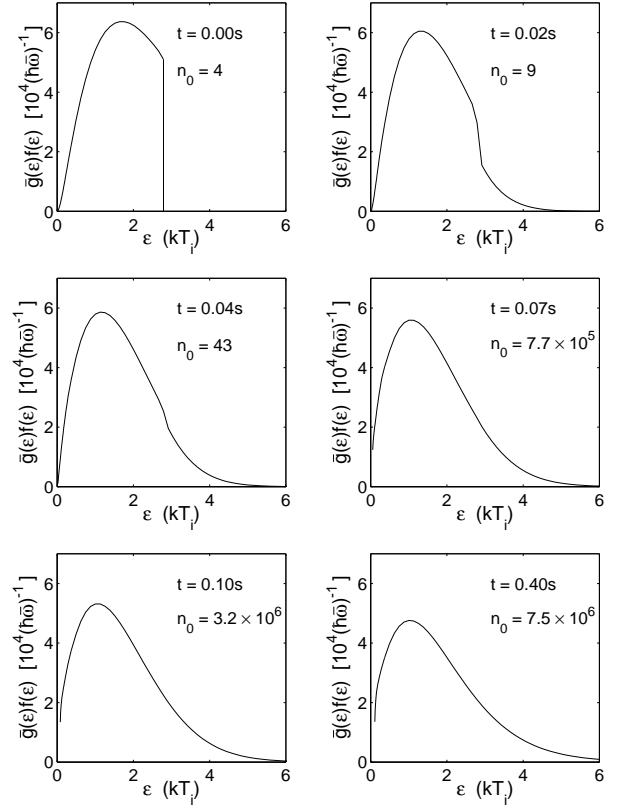,width=8cm}
\caption{Snapshots of the distribution function for a simulation with
initial conditions $\muinit=-100\hbar\bar{\omega}$, $T_i=1119$nK, and 
$\eta=2.83$.   This
results in a condensate with $n_0=7.5\times10^6$~atoms at a temperature
of $T_f=830$nK.  
For clarity, the condensate itself is not depicted, but the presence 
of a significant amount of condensate has the effect of displacing the 
left hand ends of the curves (d)--(f) an amount $\muc(n_0)/kT_i$ from the 
axis.
The growth curve for this simulation is shown in 
Fig.~\protect\ref{fig:Tf830_theory}(a).}
\label{fig:evap}
\end{figure}

We have carried out a full investigation of the effect of varying the 
initial cloud parameters has on the growth of the condensate for the 
trap configuration described in \cite{MITgrowth}.  In this paper we 
concentrate on a comparison of these results with our earlier 
theoretical model.  To model these experiments, we begin our 
simulations with an equilibrium Bose-Einstein distribution, with 
temperature $T_i$ and chemical potential $\muinit$ and truncate it at 
an energy $\varepsilon_{\rm cut} = \eta k T_i$, which represents the 
system at the end of the RF sweep.  This is then allowed to evolve in 
time, until the gas once again approaches an equilibrium the 
appropriate equilibrium Bose-Einstein in the presence of a condensate.
 This is pictured schematically in 
Fig.~\ref{fig:evap}.

\NP\NP

Because of the ergodic assumption, the MQBE that we simulate depends 
only on the geometric average of the trapping frequencies  $\bar{\omega}
= (\omega_x \omega_y \omega_z)^{1/3}$.  There is likely  to be some type
of experimental dependence on the actual trap geometry  which is not
included in our simulation; however in the regime $kT \gg \hbar 
\bar{\omega}$ this should be small.  The trap parameters of 
\cite{MITgrowth} were  
$(\omega_x,\omega_y,\omega_z)=2\pi\times(82.3, 82.3, 18)$Hz ,  
giving $\bar{\omega}=2\pi\times49.6$Hz.

\subsection{Matching the experimental data}
\label{nonlinear_eqns}

The main source of quantitative experimental data of condensate growth 
generally available is Fig.~5 of \cite{MITgrowth}.  This gives growth  rates as
a function of final condensate number and temperature rather  than the initial
conditions.  Whereas the growth curves calculated in  \cite{BosGro1,BosGro2}
required these parameters as inputs, the  calculations presented here require
three different input parameters; the  initial number of atoms in the system
$N_i$ (and hence the initial  chemical potential $\muinit$), the initial
temperature $T_i$, and the  position of the cut energy $\eta k T_i$.

\NP\begin{figure}[p]\centering
\epsfig{file=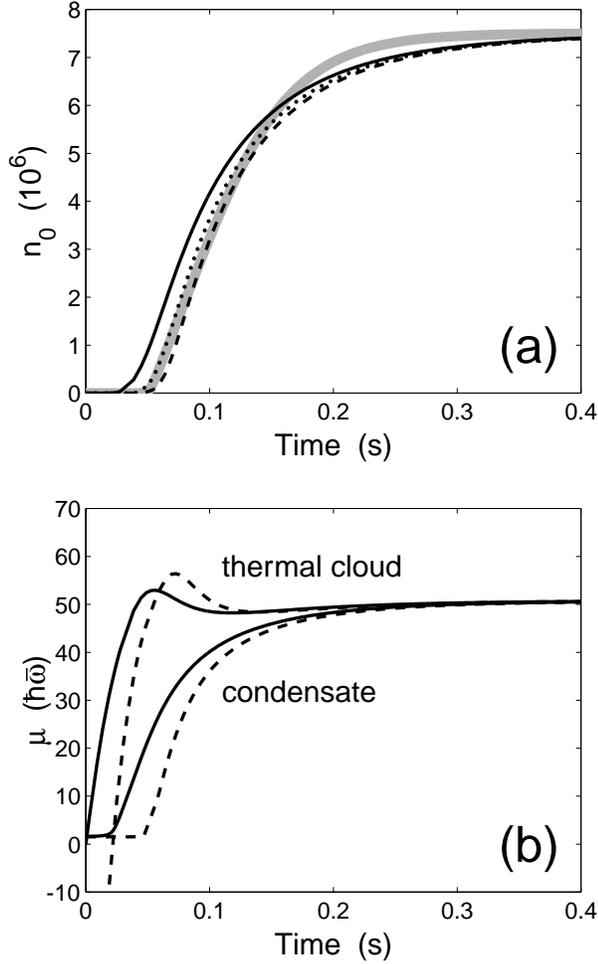,width=8cm} \caption{Growth of a 
condensate with $n_0 = 7.5\times10^6$, $T_f = 830$nK. Solid lines 
$\muinit = 0 $, dotted lines $\muinit = -40 \hbar \bar{\omega}$, 
dashed lines $\muinit = -100 \hbar \bar{\omega}.$ (a) Population of 
condensate versus time.  Grey curve is the solution for model of 
\protect\cite{BosGro2}.  (b) Chemical potential $\muc(n_0)$ of 
condensate (lower curves) and effective chemical potential $\mueff$ of 
thermal cloud (upper curves).}
\label{fig:Tf830_theory}
\end{figure}\NP

Given the final parameters supplied in \cite{MITgrowth}, it is  possible
to calculate a set of initial conditions that we require.  As  we know
the final condensate number, we can calculate the value of the  chemical
potential of the gas using the Thomas-Fermi approximation  for the
condensate eigenvalue, Eq.~\ref{eqn:mu_N}.  This gives a density of states 
according to 
Eq.~\ref{eqn:d_o_s}, and along with the measured final temperature
$T_f$, we can  calculate the total energy $E_{\rm tot}$ and number of atoms
$N_{\rm tot}$ in the system at the end of  the experiment, completely
characterizing the final state of the gas.

\begin{eqnarray}
N_{\rm tot} & =& n_0 +\sum_{\varepsilon_n > \muc(n_0)}^{\infty}
\frac{g_n}
{\exp[(\varepsilon_n- \muc(n_0))/kT_f] -1}, \\
E_{\rm tot} & = & E_0(n_0) + \sum_{\varepsilon_n > \muc(n_0)}^{\infty}
\frac{\varepsilon_n g_n}
{\exp[(\varepsilon_n- \muc(n_0))/kT_f] -1}.
\nonumber\\
\label{eqn:final_conditions}
\end{eqnarray}

\noindent We now want to find an initial distribution that would have
the same total energy and number of atoms if truncated at
$\varepsilon_{cut} = \eta k T_i$.  If we specify an initial chemical
potential for the distribution $\muinit$, we can self-consistently solve
for the parameters $T_i$ and $\eta$ from the following non-linear set of
equations
\begin{eqnarray}
N_{\rm tot} & = &\sum_{\varepsilon_n = 3\hbar\bar{\omega}/2}^{\eta k T_i}
\frac{g_n}
{\exp[(\varepsilon_n- \muinit)/k T_i] -1}, \\
E_{\rm tot} & = &\sum_{\varepsilon_n = 3\hbar\bar{\omega}/2}^{\eta k T_i}
\frac{\varepsilon_n g_n}
{\exp[(\varepsilon_n- \muinit)/k T_i] -1}.
\label{eqn:init_conditions}
\end{eqnarray}
This gives the input parameters for our simulation, and
we can now calculate growth curves starting with initially different
clouds, but resulting in the same  final condensate number and
temperature.

\subsection{Typical results}

 A sample set of growth curves is presented in Fig.~\ref{fig:Tf830_theory}(a),
for a condensate with $7.5 \times 10^6$ atoms at  a final temperature of
$830$nK and a condensate fraction of 10.4\%.  The initial parameters for the
curves are given in Table~\ref{T=830_ICs}.

As can be seen the curves are very similar, and arguably  would be difficult to
distinguish in experiment.  The main difference is the further the system
starts from the transition point (i.e. the more negative the initial chemical
potential), the longer the initiation time but the steeper the growth curve.  
This trend continues as $\muinit$ becomes more negative.

\vbox{\begin{table}\centering
\begin{tabular}{|c|c|c|c|c|}
$\muinit  (\hbar \bar{\omega})$ & $T_i  ({\rm nK})$& $N_i  (10^6)$ & $\eta $
& $ \varepsilon_{\rm cut}  (\hbar\bar{\omega})$\\
\hline
0      & 1000   & 89.1  	& 3.82  & 1605  \\
-40    & 1080   & 100.1 	& 3.31  & 1503  \\
-100   & 1119   & 117.6 	& 2.83  & 1419
\end{tabular}
\caption{Parameters for the formation of a condensate with $n_0 = 7.5 \times
10^6$ atoms at a temperature of $T_f = 830{\rm nK}$ from an uncondensed thermal
 cloud.  The growth curves are plotted in 
Fig.~\protect\ref{fig:Tf830_theory}.}
\label{T=830_ICs}
\end{table}}% 9.6% for T=590nK

\subsubsection{Effective chemical potential} 

To facilitate understanding of these results, we introduce the concept  of an
effective chemical potential $\mueff$ for the non-condensate  band.  We do this
by fitting a Bose-Einstein distribution to the  lowest energy bins of $R_{NC}$
as a function of time.  Obviously,  the chemical potential is undefined when
the system is not in  equilibrium, but as has been noted for the classical
Boltzmann  equation \cite{fitting}, the distribution function tends to
resemble  an equilibrium distribution as evaporative cooling proceeds.  The 
effective chemical potential is not unique---it is dependent on the  particular
choice of the energy cutoff $E_R$. It gives a good  indication of the ``state''
of the non-condensate, however, since the majority  of the particles entering
the condensate after a collision come from these bins. 
 In this paper $\mueff$ was computed by a
linear fit to $\log[1 +1/f_n]$ of the first ten bins of  the
noncondensate band, with the intercept giving  $\mueff$ and the gradient the
temperature.

\subsubsection{Interpretation}
We find that all the results presented in this paper can be 
qualitatively understood in terms of the simple growth equation 
Eq.~(\ref{eqn:simple_growth}), with the vapour chemical potential 
$\mu$ replaced by the effective chemical potential $\mueff$ of the 
thermal cloud.

The simple growth equation requires $\mueff > \muc(n_0) $ for  condensate
growth to occur.  In Fig.~\ref{fig:Tf830_theory}(b) we plot  the effective
chemical potential $\mueff$ of the thermal cloud and the  chemical potential of
condensate $\muc(n_0)$.  This graphs helps explain  the two  effects noted
above---longer initiation time and a steeper growth curve for the $\muinit =
-100 \hbar \bar{\omega}$ case.  Firstly, the inversion of the  chemical
potentials for this simulation occurs at a later time than  for $\muinit = 0$,
causing the stimulated growth to begin later.  This is  because the initial
cloud for the $\muinit = -100 \hbar \bar{\omega}$ simulation is further from
the transition point at $t=0$.   Secondly, the  effective chemical potential of
the thermal cloud rises more steeply,  meaning that $\mueff-\muc(n_0) $ is
larger, and therefore the rate of  condensate growth is increased.

\subsection{Comparison with earlier model}

In Fig.~\ref{fig:Tf830_theory}(a) we have also plotted the growth curve  that is
calculated for these final condensate parameters by the model of
\cite{BosGro2}, which we refer to as the simple model (not to be mistaken with
the solution of the simple growth equation \ref{eqn:simple_growth}).  For this
earlier model  the initial condensate number is indeterminate,  whereas for the
detailed calculation presented here the initial distribution is Bose-Einstein,
with the zero of the time  axis being the removal of the high energy tail.

\NP\begin{figure}\centering
\epsfig{file=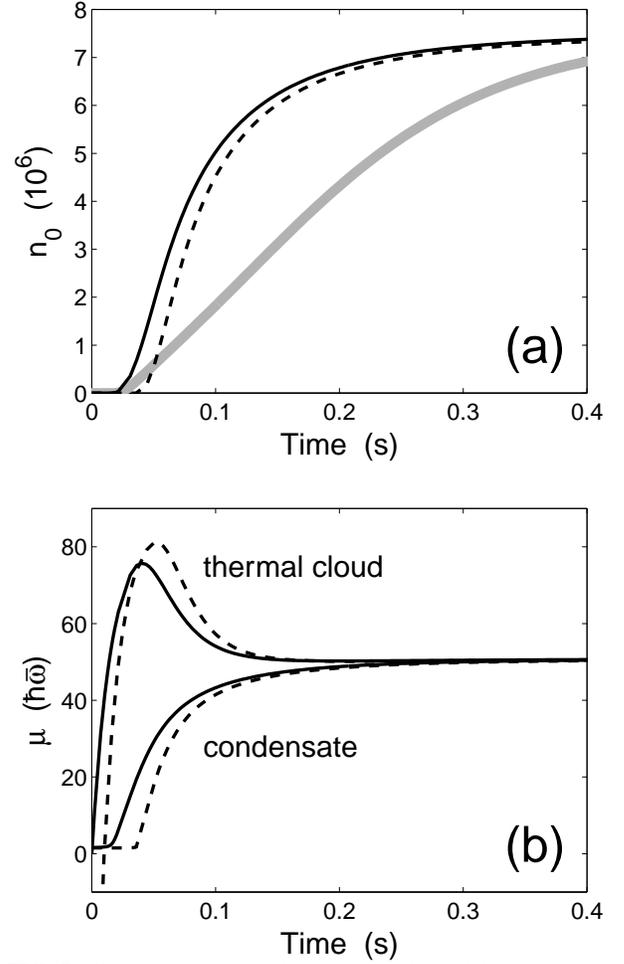,width=8cm}
\caption{Comparison of condensate growth models for a condensate fraction of
24.1\%, $n_0 = 7.5\times10^6$, $T_f = 590$nK.  
Solid lines $\muinit = 0$, dashed lines $\muinit = -100 
\hbar \bar{\omega}.$
(a) Population of condensate versus time.  Grey line is the solution for model
of \protect\cite{BosGro2}.
(b) Chemical potential of condensate (lower curves) and effective chemical
potential of thermal cloud (upper curves).}
\label{fig:Tf590_cf}
\end{figure}\NP

For these particular parameters, it turns out that the results of the 
full calculation of the growth curve give very similar results to the 
previous model, with the initial condensate number adjusted 
appropriately.  This is not surprising; indeed, from 
Fig.~\ref{fig:Tf830_theory}(b) we can see that the approximation of the 
thermal cloud by a constant chemical potential (i.e. the cloud is not 
depleted) is good for the region where the condensate becomes 
macroscopic.

For larger condensate fractions, however, the principal condition
assumed in  the model of \cite{BosGro2}, that the chemical 
potential of the vapor  can be treated as approximately constant, is
no longer satisfied.   In  Fig.~\ref{fig:Tf590_cf}(a) we plot the
growth of the same size condensate  as in
Fig.~\ref{fig:Tf830_theory},   (that is, $7.5\times10^6$ atoms) but
at a lower final temperature of 590nK.  In this situation the
condensate fraction increases to 24.1\%, and so  there is
considerable depletion of the thermal cloud.  The effect of  this
can be seen in Fig.~\ref{fig:Tf590_cf}(b).  The difference between 
the vapor and condensate chemical potentials $\mueff -\muc(n_0)$
initially increases to much larger values than for the simple model,
where $\mueff$ is held constant at its final equilibrium value. It
is this fact that causes more rapid growth. 

As the condensate continues to grow, it begins to significantly
deplete the thermal cloud, causing $\mueff$ to decrease from its maximum.   
It is the ``overshoot'' of $\mueff$ from the final equilibrium value
that the model of  \cite{BosGro2} and QKVI cannot take account of.  This
overshoot only occurs for final condensate fractions of more than
about ten percent; hence up to this value the simple model should be
sufficient.

\subsection{Effect of the final temperature on condensate 
growth}\label{temp_dep}

\begin{figure}\centering
\epsfig{file=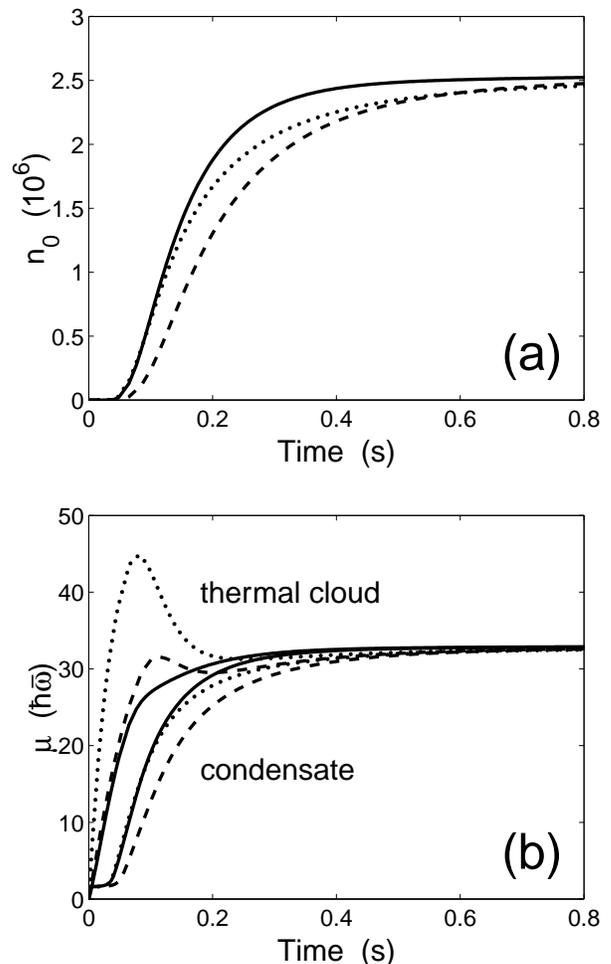,width=8cm}
\caption{Growth of a condensate with a final condensate size of
 $2.5\times 10^6$ atoms from a vapor with $\muinit = 0$.  The dotted line is for a final temperature of 400nK,
 dashed 600nK, and solid 1$\mu$K. (a) Growth curves.  (b) Chemical potential of
 condensate (lower curves) and thermal cloud (upper curves). }
\label{fig:T_depend}
\end{figure}

We have investigated the effect that final temperature has  on the
growth of a condensate of a fixed number.  All simulations were 
begun with $\muinit = 0$, since the initial chemical potential has 
little effect on the overall shape of the growth curves.  This
determines the other parameters, $T_i$ and $\eta$ and the initial 
conditions are shown in Table~\ref{tab:ICs_T}.  The results of these
simulations are  presented in Fig.~\ref{fig:T_depend}.

We find the somewhat surprising result that the growth curves do not
change significantly over a very large temperature range for the
same size condensate.  In fact, a condensate formed at 600nK grows
more slowly than at 400nK for these parameters.  As the temperature
is increased further, however, the growth rate increases again, and
at a final temperature of 1$\mu$K the growth rate is faster than at
400nK.  This effect has also been observed for both larger ($7.5
\times 10^6 $)  and smaller ($1 \times 10^6 $) condensates.

This observation can once again be interpreted using the simple
growth  equation (\ref{eqn:simple_growth}).  Although $W^+(n_0)$
increases  with temperature (approximately as $T^2$ as shown in
\cite{BosGro1}),  the maximum value of $\mueff -\muc(n_0)$ achieved
via evaporative cooling decreases with temperature for a fixed
condensate number, as the cut required is less severe and the final
condensate fraction is smaller.  Also, the term in the curly
brackets of Eq.~\ref{eqn:simple_growth} is  approximately
proportional to $T^{-1}$ for most regimes.  The end result is that
the decrease in this term compensates for the increase  in
$W^+(n_0)$, giving growth curves that are very similar for the
different simulations.  Once the  ``overshoot'' of the thermal cloud
chemical potential ceases to occur  (when the evaporative cooling
cut is not as severe), the growth rate begins to  increase with
temperature as predicted by the model of \cite{BosGro2} and QKVI.

\NP\NP

\vbox{\begin{table}
\begin{tabular}{|c|c|c|c|c|c|}
$T_f  ({\rm nK}) $& $T_i  ({\rm nK})$ &$N_i   (10^6)$ & $\eta $
 & $\varepsilon_{\rm cut}  (\hbar\bar{\omega})$& Condensate fraction\\
\hline
400	& 622.0		& 21.5	& 2.19	& 572	& 0.253	\\
600	& 707.3		& 31.6	& 4.03	& 1198	& 0.099	\\
%800	& 877.2		& 60.2	& 5.14	& 1897	& 0.046 \\
1000	&1064.8		&107.7	& 5.87	& 2629	& 0.025 	
\end{tabular}
\caption{Parameters for the formation of condensate with $n_0 = 2.5 \times
 10^6$ atoms 
from an uncondensed thermal cloud with $\muinit = 0$.  The growth curves are
presented in Fig.~\protect\ref{fig:T_depend}.}
\label{tab:ICs_T}
\end{table}}

\vbox{\NP\begin{figure}\centering
\epsfig{file=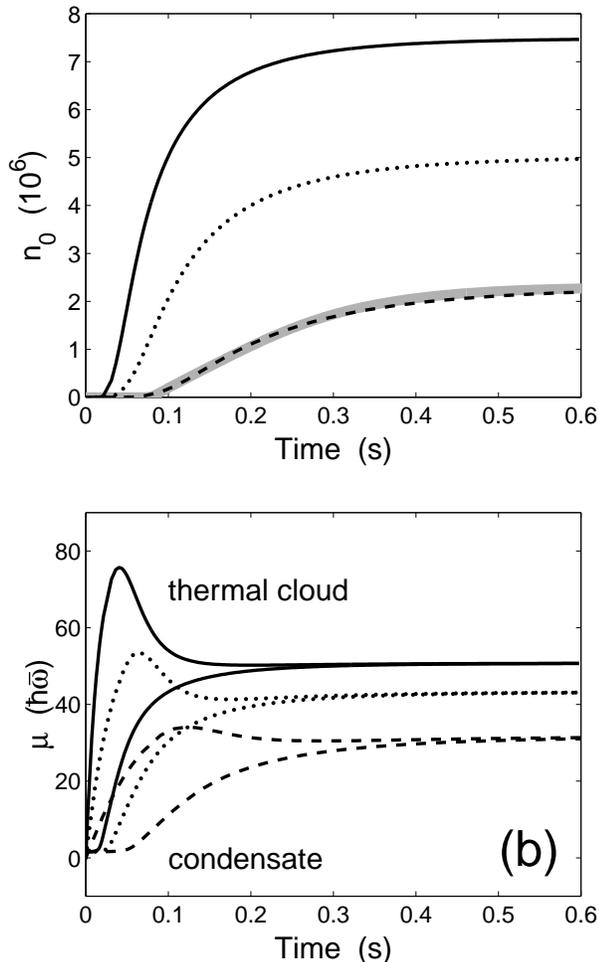,width=8cm} \caption{Growth of a condensate 
with at a final temperature of 590nK, starting from an uncondensed 
thermal cloud with $\muinit = 0$.  Solid line $7.5 \times 10^6$ 
atoms, dotted line $5.0 \times 10^6$ atoms, dashed $2.3 \times 10^6$ 
atoms.  The dashed line is for the same parameters as the lower 
temperature curves in \protect\cite{BosGro2}.  (a) Growth curves, with the
solution to the model of \protect\cite{BosGro2} in grey. (b) Chemical potential
 of condensate (lower curves) and thermal cloud (upper curves).}
\label{fig:N_depend}
\end{figure}\NP
\begin{table}
\begin{tabular}{|c|c|c|c|c|c|}
$n_0  (10^6) $& $T_i  ({\rm nK})$ &$N_i  (10^6)$ & $\eta $
 & $\varepsilon_{\rm cut}  (\hbar\bar{\omega})$& Condensate fraction\\
\hline
2.3	& 692.5		& 29.6	& 4.07	& 1186	&  0.095	\\
5.0	& 794.6		& 44.7	& 2.91	&  973	&  0.179	\\
7.5	& 897.9		& 64.6	& 2.29	&  865	&  0.239 	
\end{tabular}
\caption{Parameters for the formation of condensates at $T_f$ = 590nK
from an uncondensed thermal cloud with $\muinit = 0$.  The growth curves
are presented in Fig.~\protect\ref{fig:N_depend}}.
\label{tab:ICs_N}
\end{table}}

\subsection{Effect of size on condensate growth}

Finally we have performed some simulations of the formation of a
condensate at  a fixed final temperature, but of a varying size.  The
parameters for these simulations are given in Table~\ref{tab:ICs_N}, 
and the growth curves are  plotted in Fig.~\ref{fig:N_depend}(a).
We find that the larger the condensate, the more rapidly it grows.  
The initial clouds required to form the larger condensates not only 
start at a higher temperature (and thus have
a higher collision rate to begin with), but also they need 
to be truncated more severely, causing a larger difference in the 
chemical potentials, as seen in Fig.~\ref{fig:N_depend}(b).  Thus 
instead of these effects negating each other as in the previous 
section, here they tend to reinforce one another.  This causes the 
growth rate to be highly sensitive to the final number of atoms in the 
condensate for a fixed final temperature.  
%This could have 
%implications for the comparison with experiment, where the uncertainty 
%in the number of atoms in the condensate can be quite large (see 
%\cite{QKVI} for further discussion of this point). 

For further 
comparison with the previous model, in Fig.~\ref{fig:N_depend}(a)  the 
dashed curve is for the same parameters as for the lower temperature 
results of \cite{BosGro2}, whose prediction is plotted in grey---as 
can be seen, the two methods are in very good agreement with each 
other for this choice of parameters.  It is this particular set of final
parameters for which the discrepancy between theory and experiment remains.

\subsection{The appropriate choice of parameters}

In our computations we have taken some care to make sure that we
can  give our results as a function of the experimentally measured
{\em  final} temperature $T_f$ and condensate number $n_0$. 
Nevertheless,  it can be seen from our results that this can give
rise to  counterintuitive behavior, such as the fact that under the
condition  of a given final condensate number, the growth rate seems
to be  largely independent of temperature, because of the
cancellation noted  in Sect.~\ref{temp_dep}.  This effect has its
origin  in the quite simple fact that with a sufficiently severe cut
it is  impossible to separate the process of equilibration of the
vapor  distribution to a quasiequilibrium from the actual process of
growth  of the condensate.  In other words, the attempt to implement
the  ``ideal'' experiment in which a condensate grows from a vapor
with a  constant chemical potential and temperature cannot succeed
with a  sufficiently large cut.  Under these conditions, the
initial  temperature differs quite strongly from the final
temperature, and as  well, the number of atoms required to produce
the condensate is so  large that the vapor cannot be characterized
by a slowly varying  chemical potential during most of the growth
process.

\subsection{Comparison with experiment} \subsubsection{Comparison with  MIT
fits} The most quantitative data available from \cite{MITgrowth} is in their 
Fig.~5, in which results are presented as parameters extracted from  fits to
the simple growth equation \ref{eqn:simple_growth}.  In \cite{BosGro2} we took
two clusters of data from this figure, at the extremes of the temperature 
range for which measurements were made, and compared the theoretical  results
with the fitted experimental curves.  At the higher  temperature of 830nK the
results were in good agreement with  experiment, but at 590nK they differed
significantly, the experimental  growth rate being about three times faster
than the theoretical result.

We have performed the same calculations using the detailed model. 
The  results for 830nK are presented in Fig.~\ref{fig:Tf830_theory} 
and  those for 590nK are presented in Fig.~\ref{fig:N_depend}. 
There is a  good match between the two theoretical models at
\emph{both}  temperatures.

%%%%%%%%%%%%%%%%%%%%%%%%%%%%%%%%%%%%%%%%%%%%%%%%%%%%%%%%%%%%%%%%%%%%%%%%
%%%%%%%%%%%%%%%%%%%%%%%%%%%%%%%%%%%%%%%%%%%%%%%%%%%%%%%%%%%%%%%%%%%%%%%%
%%%%%%%%%%%%%%%%%%%%%%%%%%%%%%%%%%%%%%%%%%%%%%%%%%%%%%%%%%%%%%%%%%%%%%%%

\subsubsection{Comparison with sample growth curves}
\label{growthcurves}

In \cite{MITgrowth} some specific growth curves are also presented, and we
shall compare these with   with our computations, and those of Bijlsma {\em et
al.} \cite{Stoof}, which  appeared in preprint form after the initial
submission of this paper.

\begin{figure}\centering \epsfig{file=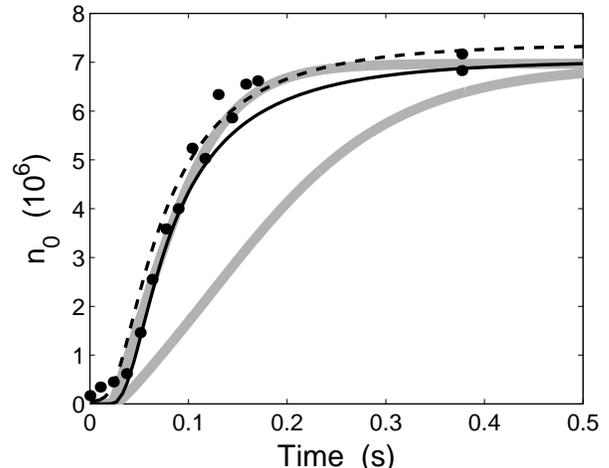,width=8cm}  
 \caption{A
comparision between the results of Fig.~11 of \protect\cite{Stoof} and our own
calculations with the initial conditions $N_i = 60 \times 10^6$  atoms, $T_i =
876$nK, $\eta = 2.5$.  Our data is  shown as the solid line and the results
from Bijlsma {\em et al.} are the dashed  line. The results of the simple model
of condensate growth with a final temperature of $T_f = 604$nK matching these
initial conditions is lower grey curve.  The upper grey curve is also for the
simple model, but with what we feel is a more realistic final temperature of
$T_f = 830$nK.  The experimental data points are the solid  dots.} 
\label{fig:stoof11} 
\end{figure}

In Fig.~\ref{fig:stoof11} we show the data from Fig.~3 of \cite{MITgrowth},
the computation of Fig. 11 of \cite{Stoof}, and our  own computations.  This is
for the MIT sodium trap, with the simulation parameters taken from
Ref.~\cite{Stoof} of $N_i = 60 \times 10^6$, $T_i =  876$nK and  $\eta = 2.5$. We find this results in a condensate
of $6.97 \times 10^6$ atoms at a temperature of 604nK, and a final condensate
fraction of 21.8\% after half a second, which agrees with our predictions from
the solution of the  equations in Sect.~\ref{nonlinear_eqns} at $t = \infty$ to
within  0.2\%.

We can see that there is little difference in the results of the two 
computations for this case, the main discrepancy being that the initiation 
time for our simulation is a little longer than that of Bijlsma {\em 
et al}.  This is likely to be due 
to the fact that their calculation starts with the condensate already 
occupied with $n_0 = 5 \times 10^4$ atoms, whereas we begin with the 
equilibrium number at this temperature given by the Bose distribution 
of $n_0 = 208$ atoms.  This difference could be brought about by the 
use of a slightly different density of states, which is also the likely 
cause of the difference in the final condensate number, of 
approximately $3 \times 10^5$ atoms.

The agreement with the experimental growth curve data is very good for
both  computations.  The simpler model of \cite{BosGro2} and QKVI cannot 
reproduce the results at this temperature, as is shown by the lower grey  curve
in Fig.~\ref{fig:stoof11}.  This is as we expect---the final  condensate
fraction is far greater than 10\% and in this case the  ``overshoot'' of
$\mueff$ is significant.

Given these initial conditions, this is the only case in which we have found
that the ``speedup'' given  by the full quantum Boltzmann theory {\em may}
yield a significant improvement  of the fit to the experimental data.

We would like to emphasise, however, that the parameters used for this
simulation {\em do not} come from Ref.~\cite{MITgrowth}. The  MIT paper does
not provide any details of the size of the thermal cloud, or the temperature at
which this curve was measured, and as such, a set of unique initial and final
parameters of the experiment cannot be determined.  We have simply taken these
parameters from the calculation of Ref.~\cite{Stoof}.

In fact, it seems likely to us that the final temperature for the experimental
curve shown in Fig.~\ref{fig:stoof11} should be higher.  Studying Fig.~5 of
Ref.~\cite{MITgrowth} shows that most condensates of $7 \times 10^6$ atoms or
more were formed at temperatures above 800nK.  We have therefore performed
a second calculation using the simple model with a final temperature of $830$nK,
and this result is shown as the upper grey curve in Fig.~\ref{fig:stoof11}. As
can be seen, {\em  this also fits the experimental data extremely well}.  The
condensate fraction at this higher temperature is 10.2\%, meaning that these
parameters are very similar to the situation considered in
Fig.~\ref{fig:Tf830_theory}, which was originally found to be a good match to
experimental data in Ref.~\cite{BosGro2}.   We note that the solution to the
simple model at this higher temperature is also in good agreement with our more
detailed calculation for these parameters.

\NP\NP

%%%%%%%%%%%%%%%%%%%%%%%%%%%%%%%%%%%%%%%%%%%%%%%%%%%%%%%%%%%%%%%%%%%%
%%%%%%%%%%%%%%%%%%%%%%%%%%%%%%%%%%%%%%%%%%%%%%%%%%%%%%%%%%%%%%%%%%%%
%%%%%%%%%%%%%%%%%%%%%%%%%%%%%%%%%%%%%%%%%%%%%%%%%%%%%%%%%%%%%%%%%%%%

The situation is very different, however, if we compare with the result of
Fig.~4 of \cite{MITgrowth},  in which the final condensate number was $1.2
\times 10^6$ atoms.
In this case, the data of Fig.~4(b) of Ref.~\cite{MITgrowth} can be used to 
extract all the relevant experimental parameters.
This graph shows an experimentally measured reduction
in the thermal cloud number from about $40 \times 10^6$ atoms to about $15
\times 10^6$ over the duration of the experiment.  Including the final
condensate population gives a total number of atoms in the system  of
approximately $16.2 \times 10^6$, or a loss of about 60\% of the atoms. 
With the three pieces of data taken from the MIT graphs (initial 
thermal cloud number, final thermal cloud number and final condensate 
number), we can estimate all the relevant parameters using the 
equations of Sect.~\ref{nonlinear_eqns}, and these are shown in the 
second column of Table~\ref{comparison}.  

\begin{figure}\centering 
\epsfig{file=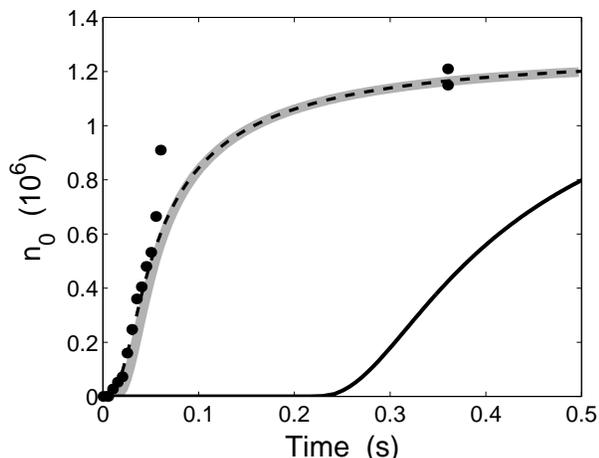,width=8cm} 
\caption{ A comparision between the data of Fig.~4 of 
\protect\cite{MITgrowth} (large solid dots)
and our own calculations. 
The solid curve shows the growth curve for the
static parameters that we have extracted from the experimental data: $N_i = 40 \times
10^6$, $T_i = 945.5$nK, $\eta = 2.19$.
An apparent fit can also be obtained---the parameters for the 
grey curve (our results) and dashed curve (Bijlsma {\em et al.}) are $N_i = 40
\times  10^6$ atoms, $T_i = 765$nK, $\eta = 0.6$.  However, as noted 
in the text, these parameters are not experimentally acceptable.   }
\label{fig:stoof10}
\end{figure}

While the parameters we present here are consistent with the static
experimental data, the growth curve corresponding to these parameters (shown in
Fig.~\ref{fig:stoof10}) certainly does not fit the dynamical data.  We find
that to remove such a large proportion of atoms, yet still obtain a relatively
small condensate, the initial system must be a long way from the transition
temperature, with $\muinit = -212 \hbar \bar{\omega}$.

This means that  condensate growth does not occur until the relaxation of the
thermal cloud is almost complete, resulting in a very long initiation time, 
Also, when the growth does begin, the rate is significantly slower than was
experimentally observed.  {\em This is the
region in which the experimental and theoretical discrepancies lie.}

\vbox{\begin{table}\centering
\begin{tabular}{|c|c|c|}
Quantity& 
\vbox{\baselineskip=9pt\hsize =2.1cm\noindent Parameters extracted
from experiment}
 & 
\vbox{\baselineskip=9pt\hsize=2.1cm\noindent
Parameters which
give an
apparent fit}
\\ \hline
$N_i$  ($10^6$)	   & 40.0        & 40.0	            \\
Atoms lost	    & 60\%      &  94\%	              \\
Condensate fraction & 7.2\%     & 51\%              \\
$T_i$  (nK)	    & 945.5    & 765	            \\
$T_f$  (nK)	    & 530      & 211	         \\
$\eta$		    & 2.19     & 0.60              \\
\end{tabular}
\caption{Comparison of the static parameters of the Bose gas that match Fig.~4
of Ref.~\protect\cite{MITgrowth} }
\label{comparison}
\end{table}}% 9.6% for T=590nK

The comparison of results is  presented in Fig.~\ref{fig:stoof10}. As well as
the computation based on the extracted parameters, we also  present  two
``apparent fits'', one based on our calculations and  another based on a
calculation of \cite{Stoof}, and here we find the  results of the two different
formulations are almost identical.  The difference appears to be due to the  
initial condensate number---our calculations begin with $295$ atoms, whereas
Bijlsma  {\em et al.} begin with $10^4$ atoms.  The initial parameters chosen
 in \cite{Stoof} for this simulation are a system of $N_i = 40 \times 10^6$ 
atoms at at temperature of $T_i = 765$nK, and the energy distribution  is
truncated at $\eta = 0.6$---an extremely severe cut.

\NP\NP

However, while the fit to the experimental data looks very good, the initial
parameters for these calculations are not consistent with the  experiment.  An
inspection of the final state of the gas  explains the situation.  The final
temperature according to these  computations is $T_f = 211$nK, and the
condensate fraction is  51\%.  Looking at the data of \cite{MITgrowth}, we find
no reported temperatures to be lower than 500nK, and the largest condensate
fraction reported to be 30\% (although our analysis of their data from Fig.~5
gave a maximum of 17\%).  The evaporative cooling of  these particular
simulations  would have to remove 94\% of the atoms in the trap, and we believe
it is very unlikely that this matches any of the experimental situations.

%%%%%%%%%%%%%%%%%%%%%%%%%%%%%%%%%%%%%%%%%%%%%%%%%%%%%%%%%%%%%%%%%%%%%%%%

\subsubsection{Speedup of condensate growth compared to simplified 
theory}

We have shown that a significant speedup of the  condensate growth can occur at
higher condensate fractions, but this   cannot explain this particular
discrepancy with the experimental results, for all of the  measured values of
temperature and condensate fraction for which growth rates are presented in
Fig.~5 of \cite{MITgrowth}.  

The only situation in which this speedup might possibly be relevant to 
experiment is the single growth curve corresponding to 
Fig.~\ref{fig:stoof11}.  However, as we have noted, the initial 
conditions for this figure are quite speculative, and in fact also 
appear to be unrealistic.

The actual speedup observed in our computations is of the
order of magnitude of that  achievable with a different condensate
fraction, and it is conceivable  that the problem could be
experimental rather than theoretical---a  systematic error in the
methodology of extracting the condensate  number from the observed
data could possibly cause the effect.  For a  realistic comparison
to be made between experiment and theory,  sufficient data should be
taken to verify positively all the relevant  parameters which have
an influence on the results.  Thus, one should  measure initial and
final temperatures, the final condensate number,  the number of
atoms in the vapor initially and finally, and the size  of the
``cut''. It should be noted in particular, in the one case where all 
of this data is available---that presented in 
Fig.~\ref{fig:stoof10}---good agreement is not found.

The previous paper in this series, QKVI, considered in detail a
semi-classical method of fitting theoretical spatial distributions
to the two dimensional data extracted by phase-contrast imaging of
the system during condensate growth.  This method  
shows that significantly different condensate
numbers and  temperatures are consistent with the MIT data and
methodology  \cite{MITgrowth}.  This seems to us to be  a more
likely origin of the  discrepancy between theory and experiment at
low temperatures with a  small condensate number.

\subsection{Outlook} 

It does remain conceivable, however, that approximations made in this 
formulation of Quantum Kinetic theory are not appropriate to the 
experimental regime where the discrepancy remains.  In this section 
we summarise the possible further extensions.

The first is the ergodic approximation, that all levels of a similar
energy are assumed to be equally occupied.  From the results of QKII
it would seem than any non-ergodicity in the initial distribution
would be damped on the time-scale of the growth---therefore the
effect of this could be significant if the initial distribution is
far from ergodic.  It is difficult to know what the exact initial
distribution of the system is without performing a three-dimensional
detailed calculation of the evaporative cooling, which would
require massive computational resources.  There is also the fact
that we have used the simplified form (\ref{ergodic7}),  derived in
analogy with the work of Holland {\em et al.}  \cite{HollandKE} on
the ergodic approximation.

The second important approximation is that the lower-lying states of
the gas are reasonably well described by the single particle
excitation spectrum, and thus using a density of states description
in calculating the collision rates of these levels.  The
justification of this is that these states are not expected to be
important in determining the growth of the condensate, and in QKVI
it was shown that varying these rates by orders of magnitude had
little effect on the growth curves.

A third approximation made is that the growth of the condensate
level  is adiabatic, and its shape remains well described by the 
Thomas-Fermi wave function.  This may not be the case, and indeed 
some collective motion during growth was observed in 
\cite{MITgrowth}.  We feel this may become important for
sufficiently  large truncations of the thermal cloud, in experiments
that could be  considered a temperature ``quench''.  Removing this
assumption would  require introducing a full description of the
lower lying  quasiparticle levels, and a time dependent
Gross-Pitaevskii equation  for the shape of the condensate.

The final approximation is that fluctuations of the occupation of the 
quantum levels are ignored.  

The agreement between the theory and the single experiment performed
so far is generally good, and there is only one regime in which there is
significant discrepancy. The removal of these approximations requires a
large amount of work, and we feel this is not justified until new 
experimental data on condensate growth becomes available.

\section{Conclusion}

We have extended the earlier models of condensate growth 
\cite{BosGro1,newerBosGro,BosGro2} and QKVI to include the full time-dependence
of both the condensate and the thermal cloud.  We have compared the results of
calculations using the full model with the simple model, and determined
that for bosonic stimulation type experiments resulting in a condensate
fraction of the order of 10\%, the model of \cite{BosGro2} and QKVI is quite
sufficient.

However, for larger condensate fractions the depletion of the thermal
cloud becomes important.  We have introduced the concept of the
effective chemical potential $\mueff$  for the thermal cloud as it
relaxes, and observed it to overshoot its final equilibrium value in
these situations, resulting in a much higher growth rate than the simple
model would predict.  Thus we have identified a mechanism for a possible
speedup that may contribute to eliminating the discrepancy with
experiment.

We have also found that the results of these calculations can be
qualitatively explained using the effective chemical potential of the
thermal cloud, $\mueff$, and the simple growth
equation (\ref{eqn:simple_growth}).  In particular, the rate of
condensate growth for the same size condensate can be remarkably
similar over a wide range of temperatures; in contrast, the rate of
growth is highly sensitive to the final condensate number at a fixed
temperature.

This model we have used in this paper eliminates all the major 
approximations in the calculation of condensate growth, apart from the 
ergodic assumption, whose removal would require massive 
computational resources.  In the absence of experimental data
sufficiently comprehensive to make possible a full comparison between 
experiment and theory, this does not at present seem justified.

In Sect.~\ref{growthcurves} we have compared the results of our 
simulations to those of Bijlsma {\em et al.} \cite{Stoof}, and found 
that our formulations are quantitatively very similar, giving growth 
curves in very good agreement with each other. The two treatments are 
based on similar, but not identical methodologies, and have been 
independently computed.  Thus the disagreement with experiment 
must be taken seriously.

\acknowledgments

MJD would like to thank Keith Burnett for his support and guidance, St. John's
College, Oxford for financial support, and Mark Lee for many useful
discussions and assistance with solutions of the simple growth model.   CWG
would also like to acknowledge fruitful discussions with Eugene  Zaremba.

This work was supported by  the Royal Society of New
Zealand under the Marsden Fund Contracts PVT-603 and PVT-902.

\EndTwoColumn
\end{document}